\documentclass[12pt]{article}
\usepackage{graphics}
\newcommand{\beq}{\begin{eqnarray}}
\newcommand{\eeq}{\end{eqnarray}}
\newcommand{\sinc}{{\rm sinc}}
\newcommand{\FR}[2]{\textstyle\frac{#1}{#2}}
\newcommand{\half}{\FR{1}{2}}
\begin{document}
\begin{center}
{\large{\bf Nonlocality -- The party may be over!}}\\ \vspace{1cm} Trevor~W.~Marshall\\
{\small{\it Dept. of Mathematics, University of Manchester,
Manchester M13 9PL, U. K.}}\vspace{1cm}
\end{center}
{\bf Abstract}. We demonstrate that the phenomenon known as
Spontaneous Parametric Down Conversion  is really an
amplification, in a nonlinear crystal pumped by a laser, of
certain pairs of modes of the electromagnetic zeropoint field. The
demonstration is achieved by showing the existence of a related
phenomenon, Spontaneous Parametric Up Conversion. This phenomenon,
once observed, will cast doubt on the quantum-optical theory,
which treats photons as the elementary objects of the light field.
It will also lend greater credibility to the zeropoint-field
description of optical entanglement phenomena. That description is
based on the unquantized light field and is consistently local, in
contrast with the nonlocal description of Quantum Optics.
\section{Introduction}
In 1948 Albert Einstein\cite{dialect} wrote a letter to Max Born
in which he enunciated the Principle of Local Action\footnote{Born
translated Einstein's {\it Nahewirkung} as `contiguity'. I think
my translation of `local action' is better; the opposite of {\it
Nahewirkung} is {\it Fernwirkung}, which is usually translated as
`action at a distance'.} (PLA), which states that any physical
action is transmitted from one point to another by a field which
propagates outwards in space and forwards in time at a speed no
greater than $c$. The usual form of Quantum Optics\cite{mandel} is
based on the notion of {\it photons} as the elementary objects of
the light field, whose corpuscular nature is revealed, through the
{\it collapse} mechanism, whenever a detection event occurs. The
PLA is violated by Quantum Optics\cite{bohm}, because a light
signal, in that description, may undergo collapse as a result of
making an observation on another signal with which it is
correlated. The nonlocal correlation between photons  has been
given the title of {\it entanglement}, in order to emphasize that,
following an analysis first made by Bell\cite{bell}, there is no
way, based on normal probability theory, in which the correlation
of the two photons may be explained in terms of their common
origin. The latter may be in the mutual annihilation of a
positron-electron pair\cite{wu}, or in an atomic
cascade\cite{claufree} or in the nonlinear optical process known
as Spontaneous Parametric Down Conversion\cite{kwiat} (SPDC).
Observations of all these  give correlations which confirm the
entanglement phenomenon.

Does this mean that Einstein was wrong, and that the PLA has to be
abandoned? Before attempting to answer that we should note that,
in his letter to Born, Einstein expressed the opinion that the PLA
is the basis of all experimental science. So mere corroboration of
the entanglement predictions is not sufficient. If anyone
seriously wishes to take on the PLA, then they must be able to
point to an experimental situation where that principle is tested
{\it without reference to any other assumptions, however plausible
those assumptions may appear to be}.

Such a necessity was recognized by Clauser and
Horne\cite{clauhorn}, who acknowledged that a local description of
an entanglement phenomenon would not necessarily be in terms of
photons. They proposed a plausible assumption which made it
possible to transform the untestable Bell inequality into a
testable Freedman inequality, namely the {\it no-enhancement
hypothesis} that  a given light signal has a certain probability
of activating a detector, and that, if a polarizer is interposed
between the source and the detector, that detection probability
cannot increase. The experiment of Freedman and Clauser showed a
violation of that inequality, and most people, including the
experimenters, came to the breathtakingly naive conclusion that
the PLA had been violated. The simple alternative of holding on to
the PLA (that is, according to Einstein, to the standard procedure
of experimental science) and concluding that a hypothesis which
had seemed very plausible was nevertheless wrong, seems not to
have occurred to anybody! Those few of us who did hold on were
occasionally acknowledged in a footnote. According to one
overworked phrase we were said to be ``exploiting the detection
loophole''\cite{gisin} in the proof of nonlocality.

We have shown\cite{pdc1,pdc2,pdc3,pdc4,pdc5,pdc6,jmo,reject}, that
entanglement in SPDC may be interpreted as a correlation between
modes of the electromagnetic field, in which both the parts above
and below the ``vacuum'', or zeropoint field (ZPF), level
participate. In this interpretation the observed enhancement
phenomena are all consistently local. We have previously
reviewed\cite{marsant,pra} the earlier generation of Bell tests
using atomic cascades, and shown that any theory of light
detectors which recognizes that they subtract the ZPF violates the
no-enhancement hypothesis.  We have now developed a full
theory\cite{reject}, at least in the context of nonlinear
crystals, of the production, propagation and detection of weak
signals. That theory now also leads to the prediction of the new
phenomenon which we call\cite{jmo,myarch} Spontaneous Parametric
Up Conversion (SPUC).

So what is ``exploiting the detection loophole''? It is simply
following the normal method of experimental science. We use
Einstein's PLA to identify the weak point of the prevailing
theory, in this case the standard quantum-optical theory of
detection. That theory is based on the normal-ordering
algorithm\cite{mandel}, which in turn comes from the  collapse
mechanism, already identified, by Einstein at the Solvay
Conference of 1927, as a weak point. Following that indication we
have been able to show that the normal-ordering algorithm may be
shown to be the linear limit of a {\it realist} detection theory,
that is one in which the activation of a detector depends only on
the local values of the electric field within a certain region of
space and within a certain time window. And that enables us to
make further experimental predictions\cite{reject}, in addition to
that of SPUC\cite{jmo}.

\section{Down conversion}
Parametric Down Conversion (PDC) is the optical analogue for a
well established classical phenomenon of wave propagation in a
nonlinear dispersive medium. It was first demonstrated in the
1960s, when high-intensity coherent sources (lasers) became
available. When two lasers, one ultraviolet (frequency $\omega_1$)
and the other infrared (frequency $\omega_2$), are incident on one
of a wide class of nonlinear crystals (NLC), then, for a certain
combination of incidence angles, a signal at the difference
frequency ($\omega_3=\omega_1-\omega_2$) is emitted (see Fig.1).
For example, if the ultraviolet wavelength is 351nm and the
infrared wavelength is 845nm, then the signal wavelength is 600nm,
which is in the visible spectrum.
\newpage
\begin{center}
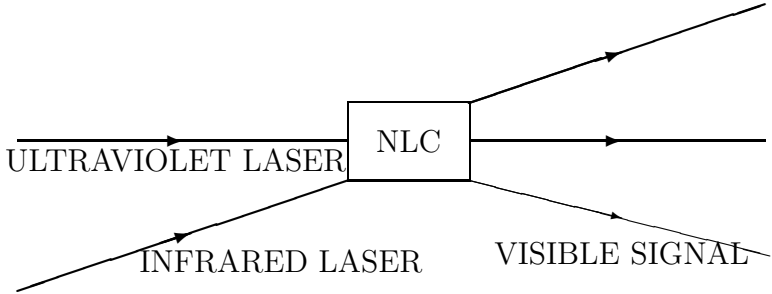
\begin{figure}[htb]
\unitlength 1mm \linethickness{0.4pt}
\begin{picture}(120.00,40.78)
\put(33,13.33){\makebox(0,0)[cc]{}}
\put(71,13.33){\makebox(0,0)[cc]{}}
\put(22,15){\makebox(0,0)[cc]{ULTRAVIOLET LASER}}
\put(45.00,12.33){\framebox(16.00,10.33)[cc]{NLC}}
\put(61.00,12.33){\line(4,-1){40.00}}
\put(61.00,12.33){\vector(4,-1){20.33}}
\put(36.00,1.67){\makebox(0,0)[cc]{INFRARED LASER}}
\put(81.33,2.67){\makebox(0,0)[cc]{VISIBLE SIGNAL}} \thicklines
\put(61.00,22.67){\vector(3,1){20.33}}
\put(61.00,22.67){\line(3,1){39.33}}
\put(20,4.00){\vector(3,1){4}}
\put(45.00,12.33){\line(-3,-1){44.00}}
\put(1.00,17.67){\line(1,0){44.00}}
\put(1.00,17.67){\vector(1,0){22.00}}
\put(61.00,17.67){\line(1,0){39.33}}
\put(61.00,17.67){\vector(1,0){20.33}}
\end{picture}
\caption{Parametric down conversion} \label{pdc}
\end{figure}
\end{center}
The angle between the two lasers, and the exit angle of the
signal, are determined through the {\it phase matching relations}
between the frequencies and wave vectors $({\bf k}_1,{\bf
k}_2,{\bf k}_3)$ of the three modes, that is \beq
\omega_3=\omega_1-\omega_2\;,\nonumber\\{\bf k}_3={\bf k}_1-{\bf
k}_2\;,\label{match}\eeq which have to be combined with the
Sellmeier relations for the refractive index of the crystal, as a
function of frequency and of the angle between the wave vector and
the polar axis. Account also has to be taken of the polarization
of the three modes; in a typical arrangement, called Type-I PDC,
$\omega_1$ is extraordinary, and $\omega_2$ and $\omega_3$ are
ordinary.

\begin{figure}[htb]
\begin{center}
\scalebox{0.7}{\includegraphics{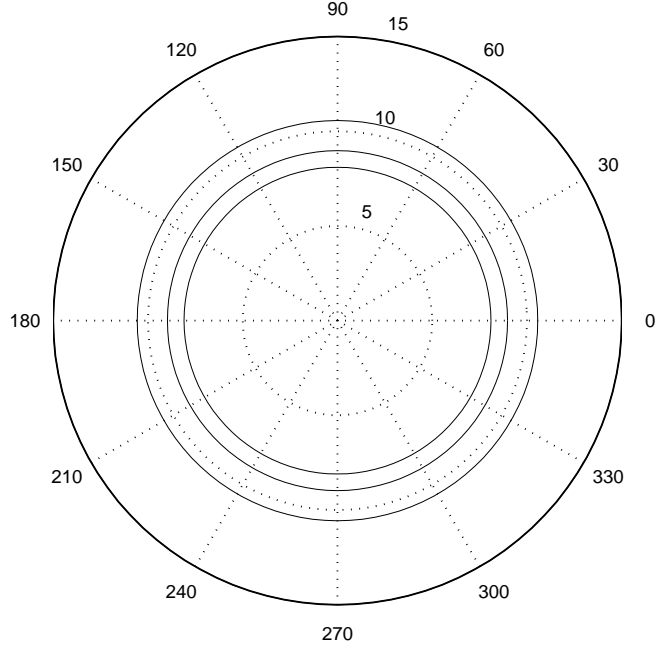}}
\end{center}
\caption{Position of the SPDC rainbow produced when a 351nm laser
is normally incident on a BBO crystal cut with its axis at 37
degrees to the incident wave vector. The 600, 700 and 800nm
components are shown; the 600 being the inner one and the 800 the
outer one. } \label{pdcrain}
\end{figure}

Spontaneous Parametric Down Conversion (SPDC) is the name given to
the phenomenon which occurs when we remove the laser $\omega_2$; a
weak signal, which is nevertheless visible to the unaided eye,
remains in the  $\omega_3$ channel. This is because, in the
vacuum, there is a {\it zeropoint field} (ZPF) intensity in all
modes, corresponding to half a ``photon". The radiation pressure
of this ZPF field is what gives rise to the Casimir
effect\cite{mex,milonni}. Through its nonlinear interaction with
the hydrogen atom, the ZPF also gives rise to the greater (that is
nonrelativistic) part of the Lamb shift in the hydrogen
spectrum\cite{milonni}. In SPDC, as opposed to PDC, the laser mode
${\bf k}_2$ is replaced by the same, but very much less intense,
ZPF mode, so ${\bf k}_3$ is still emitted. The SPDC phenomenon
actually manifests itself as a rainbow (see Fig.\ref{pdcrain}).
Since all frequencies and directions of modes are present in the
ZPF, all the corresponding down converted signals also appear at
those angles satisfying the phase matching relations.

The above phenomenon is nowadays often called simply PDC, because
it has become popular to view the process as one in which laser
photons $\omega_1$ down convert into $\omega_2$ and $\omega_3$. In
this description the phase matching relations linking the wave
vectors of the three coupled modes in SPDC are considered to
express the conservation of four-momentum between the three
participating photons, that is \beq
\hbar\omega_1=\hbar\omega_2+\hbar\omega_3\;,\nonumber\\\hbar{\bf
k}_1=\hbar{\bf k}_2+\hbar{\bf k}_3\;.\label{conserve}\eeq These
equations are, of course, trivially equivalent to (\ref{match}),
but their theoretical underpinning is entirely different. We may
write eq.(\ref{match}) in the descriptive form
\begin{eqnarray*}
{\rm laser\; ultraviolet\; wave+vacuum\; infrared\;wave\rightarrow
visible \;signal}\;,
\end{eqnarray*}
whereas the corresponding description for eq.(\ref{conserve}) is
\begin{eqnarray*}
{\rm laser\; ultraviolet\;photon \rightarrow visible
\;signal+infrared\;idler}\;.
\end{eqnarray*}
It is an unfortunate and highly misleading consequence of this
latter description that certain correlations in the intensities of
the outgoing SPDC signals are now widely interpreted as showing
rather bizarre connections between the corresponding photons. All
of these allegedly nonlocal
phenomena\cite{kwiat,hong,shihal,wzm,shih,weihs}, including those
which purport to test the Bell inequalities\cite{kwiat,shihal},
have the entanglement feature in common, and our analysis, based
on the wave description of light, gives a consistently local
description for all of them. The difference is, perhaps,
emphasized by the absence of $\hbar$ in eq.(\ref{match}), as
opposed to its presence in (\ref{conserve}). The optical theory we
are proposing is a {\it semiclassical} one, that is it combines a
wave description of the light field with a quantum description of
atoms. However, the interaction of light with a nonlinear crystal
does not require any recognition of the atomic constitution of the
crystal. Consequently our description of such phenomena, in
contrast with our description of atomic cascades, is {\it purely
classical}. Planck's constant enters only when we want to
calculate the {\it intensity} of the SPDC rainbow, which we shall
do in the next section. That is because $\hbar$ is the constant
determining the intensity of the ZPF. I think Max Planck would
have approved, because he originated the idea of the ZPF in
1911\cite{mex,milonni,planckz}.
\section{Up conversion}
The SPUC phenomenon, which was mentioned in the Introduction, is a
simple modification of SPDC. It has not yet been observed, but any
well equipped optical laboratory with a suitably cut specimen of
the nonlinear  crystal  beta barium borate (BBO) may see it by
following the design details below. Essentially all that needs to
be done is to return to Fig.\ref{pdc} and, instead of removing the
infrared laser (that is $\omega_2$), we remove the ultraviolet one
(that is $\omega_1$). Then the signal $\omega_3$ persists. We may
describe this process as
\begin{eqnarray*}
{\rm laser\; infrared\; wave+vacuum\; ultraviolet\;wave\rightarrow
visible \;signal}\;,
\end{eqnarray*}
but, in contrast with the corresponding SPDC process, there is no
photonic equivalent. However, the phase matching relations for the
SPDC and SPUC processes are identical. Therefore we can be assured
that, {\it if one of them is a real process, then so is the
other}. It only remains to calculate the position and the
intensity of the SPUC rainbow.

\begin{figure}[htb]
\begin{center}
\scalebox{0.7}{\includegraphics{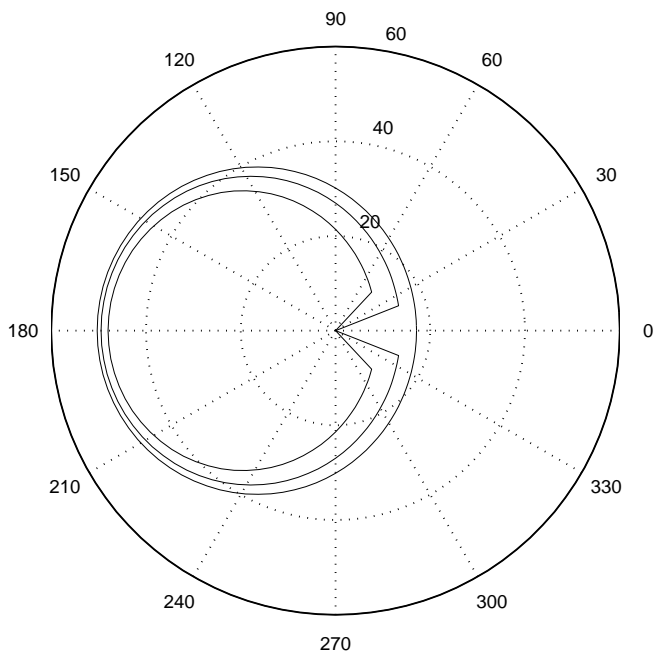}}
\end{center}
\caption{Position of the SPUC rainbow produced when a 845nm laser
is normally incident on a BBO crystal cut with its axis at 37
degrees to the incident wave vector. The arcs of the 600, 700 and
800nm components are shown. Note that, in contrast with SPDC
(Fig.\ref{pdcrain}), the rainbow is not a complete circle, neither
is it centred around the direction of the pumping laser. }
\label{pucrain}
\end{figure}

For definiteness we assume that the crystal is cut with its optic
axis at 37 degrees to the incident face, and that the infrared
laser has a wavelength of 845nm normally incident. Because the
ultraviolet mode is extraordinarily polarized, its refractive
index depends on its orientation, which results in the SPUC
rainbow being off centred with respect to the laser beam. We
depict the position of the rainbow in  Fig.\ref{pucrain}.
We note that  phase matching is not possible for all azimuthal
angles, the polar axis having azimuth $\phi=0$. The angles at
which the visible SPUC modes exit from the crystal are
substantially larger than those of the SPDC rainbow, depicted in
Fig.\ref{pdcrain}, especially since, as we shall see shortly, the
most intense part of the rainbow is near $\phi=180$ degrees, for
which $\theta$ is around 50 degrees. It is therefore not
surprising that the SPUC rainbow has never been accidentally
observed.

\begin{center}
\begin{figure}[htb]
\scalebox{0.75}{\includegraphics{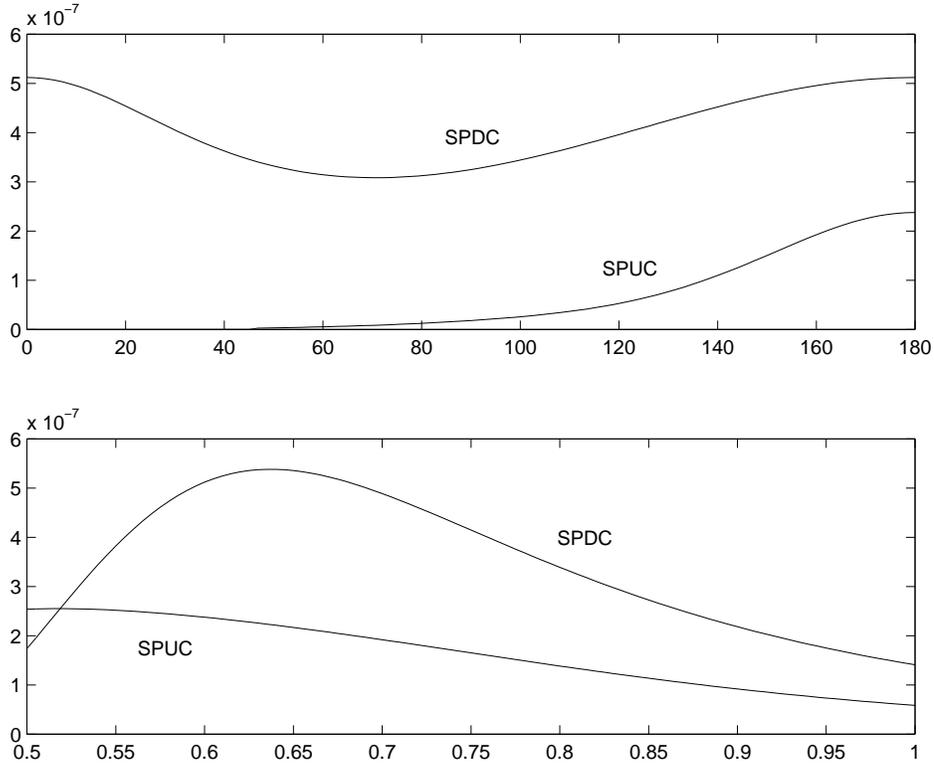}} \caption{The SPUC
and SPDC cross sections plotted against the azimuthal angle in
degrees (upper figure, outgoing wavelength .6$\mu$m) and
wavelength in $\mu$m(lower figure, outgoing azimuth 180 degrees)
for a BBO crystal cut with its optic axis at 37 degrees to the
normal of the incident face. The normally incident lasers have
wavelengths .442 $\mu$m (SPDC) and .845 $\mu$m (SPUC)}
\label{spucazi}
\end{figure}
\end{center}

The intensities of the  SPDC and SPUC rainbows are calculated from
the mode coupling equations\cite{jmo}. We make the linearization
approximation, which consists of treating the amplitude  of the
laser, that is $A_1$ in SPDC and $A_2$ in SPUC,  as a constant.
Then, for SPUC, taking  the normal to the incident face of the
crystal as the $z$-direction, $A_1$ and $A_3$ are slowly varying
functions of $z$ satisfying the coupling equations
\begin{eqnarray}
\frac{dA_1}{dz}&=&i\beta_1A_2e^{i\Delta z}A_3\;,  \nonumber \\
\frac{dA_3}{dz}&=&i\beta_3A_2^*e^{-i\Delta z}A_1\;,
\end{eqnarray}
where our model assumes perfect phase matching in the $x$ and $y$
directions, and $\Delta$ is the mismatch in the $z$ components of
${\bf k}_1,{\bf k}_2,{\bf k}_3$, that is
\begin{eqnarray}
\Delta=k_{1z}-k_{2z}-k_{3z}\;,
\end{eqnarray}
and $\beta_1,\beta_3$ are constants related to the Pockels
coefficients of the crystal. Putting the single-mode intensities
$I_r=\vert A_r\vert^2\;(r=1,2,3)$, then, for a crystal of depth
$l$, the relation between incoming and outgoing intensities is
\beq
I_1(l)-I_1(0)&=&\beta_3I_2l^2\;[\beta_3I_1(0)-\beta_1I_3(0)]\;\sinc^2(\half\Delta'
l)
\;,\nonumber\\I_3(l)-I_3(0)&=&\beta_1I_2l^2\;[\beta_1I_3(0)-\beta_3I_1(0)]\;
\sinc^2(\half\Delta' l)\;,\eeq where \beq \sinc(x)=\frac{\sin
x}{x}\quad,\quad \Delta'=\sqrt{\Delta^2+\beta_1I_2\beta_3}\;.\eeq
The total intensity in the $\omega_3$ channel is obtained by
summing $I_3(l)-I_3(0)$ over all the relevant pairs of modes. For
details refer to Ref.\cite{myarch}. The results are depicted in
Fig.\ref{spucazi}. Over a large part of the visible frequency
range, the SPUC intensity is about half the SPDC intensity, based
on a pumping laser for the latter at 442nm, having the same
intensity as the SPUC pump at 845nm. So, with a suitable camera,
the SPUC rainbow will be seen in glorious technicolor!

\end{document}